\documentclass[aps,prl,twocolumn,groupedaddress,showpacs]{revtex4-1}

\usepackage{bm}
\usepackage{amsthm, amsmath, amssymb}
\usepackage{latexsym,amsmath,amsfonts,amsthm,amssymb}
\usepackage{graphics}

\usepackage[breaklinks=false,pdfpagemode=None,pdfview=FitH,pdfstartview=FitH,citebordercolor={0 0 1},linkbordercolor={0 0 1},urlbordercolor={0 0 1},pagebordercolor={0 0 1},pdfborder={0 0 1}]{hyperref}

\newcommand{\veps}{\varepsilon}

\newtheorem{theorem}{Theorem}

\begin{document}


\title{Diagonalization and Many-Body Localization for a Disordered Quantum Spin Chain}


\author{John Z. Imbrie}
\affiliation{Department of Mathematics, University of Virginia,
Charlottesville, VA 22904-4137, USA}



\begin{abstract}
We consider a weakly interacting quantum spin chain with random local interactions. 
We prove that many-body localization follows from a physically reasonable assumption that limits the extent of level attraction in the statistics of eigenvalues. In a KAM-style construction, a sequence of local unitary transformations is used to diagonalize the Hamiltonian by deforming the initial tensor-product basis into a complete set of exact many-body eigenfunctions. 
\end{abstract}


\maketitle


In the past few years, there has been a surge of interest in the phenomenon of many-body localization (MBL). In the well-studied Anderson tight-binding model \cite{Anderson1958} a particle moving in a sufficiently strong random potential is localized; eigenstates decay exponentially away from localization centers and transport is absent. A number of authors have argued that localization persists in the presence of weak inter-particle interactions \cite{Fleishman1980,Giamarchi1987,Altshuler1997,Santos2004,Gornyi2005,Basko2006,Znidaric2008}. In particular, the detailed perturbative analysis of \cite{Basko2006} provided strong evidence for MBL. More recently, numerical studies on one-dimensional spin systems and particle systems \cite{Oganesyan2007,Pal2010,Bauer2013} gave evidence for a transition from a thermalized phase to a many-body localized phase, as the strength of the disorder increases. See \cite{Nandkishore2015} for a review of recent work on MBL.

On a theoretical level, it is important to get past perturbative analysis, as rare regions with weak disorder (Griffiths regions \cite{Griffiths1969}) have the potential to spoil localization. Rigorous results on localization in many-body systems include a proof of dynamical localization for an isotropic random spin chain, using the Jordan-Wigner transformation to reduce the problem to an equivalent one-body Hamiltonian \cite{Hamza2012a}. Localization in the ground state of the interacting Aubry-Andr\'e model was established in \cite{Mastropietro2015}.

In this Letter we establish rigorously that for a one-dimensional disordered spin chain, MBL follows from a physically reasonable assumption on level statistics. We consider a random field, random transverse field, random exchange Ising model on an interval
 $\Lambda = [-K,K'] \cap \mathbb{Z}$:
\begin{equation}\label{(1)}
H = \sum_{i=-K}^{K'} h_i S_i^\text{z} + \sum\limits_{i=-K}^{K'} \gamma_i S_i^\text{x} + \sum\limits_{i=-K-1}^{K'} J_i S_i^\text{z} S^\text{z}_{i + 1}.
\end{equation}
Here $S_i^{\text{x},\text{z}}$ are Pauli matrices, with $S_i^{\text{z}} \equiv 1$ for $i \notin \Lambda$.
We take $\gamma_i = \gamma\Gamma_i$ with $\gamma$ small. 
Thus the Hamiltonian is close to one that is diagonal in the basis given by tensor products of $S_i^{\text{z}}$ eigenstates.
We take the random variables $h_i, \Gamma_i, J_i$ to be independent and bounded, with bounded probability densities.
This is a variant of the model considered in \cite{Pal2010}. We will need to make an assumption of ``limited level attraction'' for the spectrum of $H$, for some values of $\nu > 0$ and $C < \infty$:

\noindent
\textbf{Assumption} \textbf{LLA}$(\nu,C)$. \textit{Consider the Hamiltonian $H$ in boxes of size $n$. Its eigenvalues satisfy
\begin{equation}\label{(2)}
P\left(\min_{\alpha \ne \beta} |E_{\alpha} - E_{\beta}| < \delta\right) \leq \delta^{\nu} C^n, 
\end{equation}
for all $\delta>0$ and all n.}

Physically, this is a mild assumption specifying that with high probability the minimum level spacing should be no smaller than some exponential in the volume. Note that random matrices normally have either neutral statistics ($\nu = 1$, \textit{e.g.} Poisson) or repulsive ones ($\nu > 1$, \textit{e.g.} GOE). Our analysis works even for attractive statistics, \textit{i.e.} $0 < \nu < 1$.
Mathematically, techniques to prove estimates such as (\ref{(2)}) are not yet available for many-body systems, but a promising approach is available for single-body Hamiltonians \cite{Imbrie2016a}.

We give an explicit construction of a sequence of unitary rotations that diagonalizes the Hamiltonian. Each rotation is generated by quasi-local operators. This means that a rotation generator that involves $\ell$ spins is exponentially small in $\ell$, with high probability. Resonant regions where the required rotations are far from the identity are dilute; the probability that two sites a distance $D$ apart are in the same resonant region decays faster than any power of $D$. These rotations define a way to deform the original basis states (tensor products of (1,0) or (0,1) at each site) into the exact eigenstates. Away from resonant regions, each eigenstate resembles the basis state it came from, and classical spin configurations $\sigma = \{\sigma_i\} \in \{-1,1\}^{|\Lambda|}$ can be used as eigenstate labels. This is made evident by the following result:
\begin{theorem} \label{thm:1}
Let $\nu, C$ be fixed. There exists a $\kappa > 0$ such that for $\gamma$ sufficiently small,
 \textbf{LLA($\nu$,$C$)} implies the following estimates:
 \begin{equation}\label{(1.4)}
\mathbb{E}\, \normalfont{\text{Av}}_\alpha \left| \langle S^\text{z}_0 \rangle_\alpha\right| = 1 - O(\gamma^\kappa), 
\end{equation}
where $\mathbb{E}$ denotes the disorder average, $\mathrm{Av}_{\alpha}$ denotes an average over $\alpha$, and $\langle \cdot \rangle_\alpha$ denotes the expectation in the eigenstate $\alpha$. For any $i \ne j$, 
\begin{equation}\label{(4)}
\max_\alpha |\langle\mathcal{O}_i;\mathcal{O}_j\rangle_\alpha| \le \gamma^{|i-j|/3}
\end{equation}
with probability
$1-(\gamma^\kappa)^{1+(\log|i-j|)^2}$. Here $\langle\mathcal{O}_i;\mathcal{O}_j\rangle_\alpha \equiv \langle\mathcal{O}_i\mathcal{O}_j\rangle_\alpha - \langle\mathcal{O}_i\rangle_\alpha\langle\mathcal{O}_j\rangle_\alpha$, with $\mathcal{O}_i$ any operator formed from products of $S^\mathrm{x}_{i'}$ or $S^\mathrm{z}_{i'}$, for $i'$ near $i$. All bounds are uniform in $\Lambda$.
\end{theorem}
We may take $\text{Av}_\alpha$ to be any normalized average over the $2^n$ values of $\alpha$ (for a box of size $n$), \textit{e.g.} uniform weights (infinite temperature) or $\text{(const)}\exp(-\beta E_\alpha)$.
We see that with high probability, most states have the property that the expectation of $S_0^{\text{z}}$ is close to $+1$ or $-1$, just like the basis vectors. The spins are effectively frozen in place in each eigenstate. This would contrast with a thermalized phase, wherein states resemble thermal ensembles (a consequence of the eigenstate thermalization hypothesis \cite{Deutsch1991,Srednicki1994,Rigol2008}). At infinite temperature, thermalization would imply that eigenstate expectations of $S_0^{\text{z}}$ would go to zero in the infinite volume limit. Thus the bound (\ref{(1.4)}) implies a failure of thermalization, a key feature of the MBL phase.

In the course of the proof, we construct a sequence of rotations (or changes of basis) that give an explicit quasi-local deformation between basis vectors and the exact eigenstates. This is an important feature of the fully many-body localized phase \cite{Bauer2013,Imbrie2016}; the property is essentially equivalent to the existence of a complete set of quasi-local integrals of motion \cite{Huse2014,Serbyn2013,Ros2015}. It is the many-body analog of the fact that one can deform (or label) single-particle eigenstates by their localization centers \cite{Imbrie2015}. 
 
\textit{Proof of Theorem \ref{thm:1}.} We present the key elements of the proof; more technical aspects are published separately \cite{Imbrie2016}.
We obtain a complete diagonalization of $H$ by successively eliminating low-order off-diagonal terms as in Newton's method. The process runs on a sequence of length scales $L_k = (\frac{15}{8})^k$, with off-diagonal elements of order $\gamma^m$, $m \in [L_k, L_{k+1})$  eliminated in the $k^{\text{th}}$ step.
The orthogonal rotations that accomplish this can be written as a convergent graphical expansion, provided nonresonant conditions are satisfied. Resonant regions are diagonalized as blocks in quasi-degenerate perturbation theory. As in the Kolmogorov-Arnold-Moser (KAM) theorem constructing integrals of motion in Hamiltonian dynamical systems, one needs to control the measure of resonant sets where perturbation arguments break down.
See \cite{Imbrie2015} for an application of these ideas to the Anderson model. 
KAM ideas have been useful in quantum models with quasiperiodic potentials \cite{Bellissard1983b,Chulaevsky1991,Eliasson1997,Mastropietro2015}.

We begin by outlining the first step of the procedure ($k=1$).

\textit{Resonances.}
Perturbation theory works if there are gaps between eigenvalues. Initially, the only off-diagonal term is $\gamma_i S^\text{x}_i$, which is local. We only need to worry about single-flip resonances (for the moment). 
Let the spin configuration $\sigma^{(i)}$ be equal to $\sigma$ with the spin at $i$ flipped. The associated change in energy is
\begin{equation}\label{(5)}
\Delta E_i \equiv E(\sigma) - E (\sigma^{(i)}) = 2 \sigma_i (h_i + J_i \sigma_{i + 1} + J_{i - 1} \sigma_{i - 1}). 
\end{equation}
We say that the site $i$ is resonant if $|\Delta E_i|<\varepsilon \equiv \gamma^{1/20}$ for at least one choice of $\sigma_{i-1}, \sigma_{i+1}$.
Then for nonresonant sites the ratio $\gamma_i/ \Delta E_i$ is  $ \le \gamma^{19/20}$.
A site is resonant with probability $\sim 4\varepsilon$. Hence resonant sites form a dilute set  where perturbation theory breaks down.

\textit{Perturbation Theory.}
Let $H=H_0 + \mathcal{J}$ with $H_0$ diagonal and $\mathcal{J}$ off-diagonal. Put
$
\mathcal{J}=J^{\textrm{res}}+J^{\textrm{per}},
$
where $J^{\textrm{res}}$ contains terms $J(i) \equiv \gamma_i S_i^{\text{x}}$ with $i$ resonant, and $J^{\textrm{per}}$ contains the rest. Then define an antisymmetric matrix
\begin{equation}\label{(6)}
A \equiv \sum_{\text{nonresonant }i} A(i) \text{ with }A(i)_{\sigma \sigma^{(i)}} = \frac{J(i)_{\sigma \sigma^{(i)}}}{E_\sigma-E_{\sigma^{(i)}}}.
\end{equation}
First-order perturbation theory can be implemented by using $\Omega = e^{-A}$ for a basis change.
This leads to a
renormalized Hamiltonian:
\begin{equation}\label{(7)}
H^{(1)}=e^{-A} H e^A.
\end{equation}
By construction (first-order perturbation theory), 
\begin{equation}\label{(8)}
[A,H_0] = -J^{\rm per},
\end{equation}
which cancels all but the resonant terms to leading order:
\begin{align}\label{(9)}
H^{(1)}
 &=  e^AHe^{-A} = H+[A,H]+ \frac{[A,[A,H]]}{2!} + \ldots 
  \nonumber
\\& =  H_0+ J^{\rm res}+\sum_{n=1}^\infty\frac{(\text{ad}\,A)^n}{n!}
\left(\frac{n}{n+1}J^{\rm per}+J^{\rm res}\right)
\nonumber
\\& \equiv  H_0+ J^{\rm res}+J^{(1)},
\end{align}
where $(\text{ad}\,A)B\equiv [A,B]$. A similar transformation was used in \cite{Datta1996}.

\textit{Properties of the new Hamiltonian:}
After the change of basis,
$J^{\rm per}$ is gone, while
$J^{\rm res}$ remains.
The new interaction
$J^{(1)}$ is quadratic and higher order in $\gamma$.
Note that
$A(i)$ commutes with $A(j)$ or $J(j)$ if $|i-j|>1$.
Thus we preserve quasi-locality of $J^{(1)}$; it can be written as 
$\sum_g J^{(1)}(g)$, where $g$ is
a sum of connected graphs involving spin flips $J(i)$ and associated energy denominators as in (\ref{(6)}). 
Specifically, a graph is determined by a sequence of sites $i_0,\ldots,i_n$ such that $\text{dist}(i_p,\{i_0,\ldots,\i_{p-1}\}) \le 1$ for $1\le  p \le n$; this specifies a nonvanishing term in $(\text{ad}\,A)^n J$ that operates on the spins at those sites.
A graph involving $\ell$ spin flips has $\ell-1$ energy denominators and is bounded by $\gamma(\gamma/\veps)^{\ell-1}$.
    
We define resonant blocks by taking connected components of the set of sites belonging to resonant graphs. We perform exact rotations $O$ in small, isolated resonant blocks to diagonalize the Hamiltonian there. This paves the way for reintegrating such regions into the perturbative framework in subsequent steps.

\textit{Expectations in approximate eigenstates.}
$S^\text{z}_0$ is diagonal in the $\sigma$-basis, so $(S^\text{z}_0)_{\sigma \tilde{\sigma}} = \sigma_0 \delta_{\sigma_0 \tilde{\sigma}_0}$. We prove a version of (4) for the eigenfunctions as approximated so far (that is, for the columns of $\Omega O$):
\begin{equation}\label{(10)}
\mathbb{E} \; \text{Av}_{\alpha}\left|\Big| \sum_{\sigma} (O^{\text{tr}} \Omega^{\text{tr}})_{\alpha\sigma} \sigma_0 (\Omega O)_{\sigma\alpha}\Big|-1\right| \le O(\varepsilon). 
\end{equation}
To handle resonances, one may throw out a set of small probability $O(\varepsilon)$ where the origin is in a resonant block and the rotation $O$ acts nontrivially. Then the rotation $\Omega$ is close to the identity.
Hence the two terms cancel except for interactions involving $A(i)$, which is bounded by $\gamma/\varepsilon$.
To obtain Theorem \ref{thm:1}, we need to prove this for the complete diagonalization of $H$.  
Similar methods should lead to a uniform bound on the entanglement entropy of a subsystem, because the only contributions come from rotations straddling the boundary.

\textit{Multi-scale iteration:} We continue the process on a sequence of length scales $L_k =(\frac{15}{8})^k$, so that off-diagonal elements of $H^{(k)}$ will be of the order $\gamma^{L_k}$. In each step, the diagonal elements of $H^{(k)}$ are renormalized by interactions up to the $k^{\rm th}$ scale; they are denoted $E^{(k)}_\sigma$.
We say that
 $g$ is resonant if 
 \begin{equation}\label{(10a)}
 A^{(k+1)}_{\sigma \tilde{\sigma}}(g) \equiv \frac{J^{(k)}_{xy}(g)}{E^{(k)}_\sigma-E^{(k)}_{\tilde{\sigma}}}  
 \end{equation}
is greater than $(\gamma/\varepsilon)^{|g|}$ in magnitude, where
$|g|$ denotes the number of spin flips in $g$;
we have $|g| \ge L_k$ in the $k^{\text{th}}$ step (as in Newton's method, the procedure has a  convergence rate that is close to quadratic).
For nonresonant graphs, we may use (\ref{(10a)}) to generate the next rotation.
Then the next interaction $J^{(k+1)}$ is again given by a sum of graphs 
$\sum_g J^{(k+1)}(g)$. This is a recursive construction; repeated application of the ad expansion (\ref{(9)}) leads to graphs $g$ that are given by a sequence of graphs from the previous scale.
The precise definition of $g$ is somewhat involved (see Appendix 1 of \cite{Imbrie2016}), but as in the first step, we may use the condition of nonvanishing commutators to enforce connectivity. Then if one unwraps
the expansions, one obtains a sequence of spin flips at a set of sites that is nearest-neighbor connected. 
Each subgraph from an earlier scale comes with an energy denominator representing the result of flipping the spins of the subgraph; there are a total of
 $|g|-1$ energy denominators.

Let us discuss a key estimate giving control over the probability of resonances. For the moment we make some simplifying assumptions: (a) The graph $g$ involves spin flips in regions not previously found to be resonant;
(b) $g$ does not flip any spin more than once;
(c) the ad expansion (\ref{(9)}) is developed only to up to order $n_0$ on each scale (this avoids complications with factorials in $n$). 

We need to maintain uniform exponential decay on the probability that $g$ is resonant. 
One cannot simply bound denominators from below, as was done in the first step. There is a mixture of energy denominators from different scales, and combining lower bounds 
$O(\veps^{L_j})$ for denominators on scales $j \le k$ would cause the rate of decay for graphs to degenerate as $k$ increases. Instead, we prove that each graph $g$ obeys a fractional moment bound with $s=\frac{2}{7}$ when averaged over the disorder:
\begin{equation}\label{(11)}
\mathbb{E}\,|A^{(k)}_{\sigma\tilde{\sigma}}(g)|^s  \le \gamma^{s|g|} \mathbb{E} \prod_{\tau \tilde{\tau}\in g} \left|E^{(j)}_{\tau} - E^{(j)}_{\tilde{\tau}}\right|^{-s} \le (c\gamma)^{s|g|}.
\end{equation}
Here the constant $c$ does not depend on $k$.
By (a), energy denominators $E^{(j)}_{\tau} - E^{(j)}_{\tilde{\tau}}$ are given by a sum of $\pm 2h_i$ over the sites $i$ flipped in the transition $\tau \rightarrow \tilde{\tau}$ (up to corrections of order $\gamma$). 
By (b), each energy denominator contains an independent integration variable $h_i$. Boundedness of the density for $h_i$ implies that $\mathbb{E}|\Delta E|^{-s}$ is bounded, and (\ref{(11)}) follows.
Then Markov's inequality implies that
\begin{equation}\label{(12)}
P\left(|A^{(k)}_{\sigma\tilde{\sigma}}(g)| > (\gamma/\veps)^{|g|}\right)
\le (c\veps)^{s|g|}.
\end{equation}
This procedure estimates all of the denominators of $g$ together; it yields the desired exponential bound on the rotation generator with probability $1-(c\veps)^{s|g|}$.

The set of sites that belong to resonant graphs $g$ are decomposed into connected components. The result is the set of resonant blocks in the $k^{\text{th}}$ step.
 The number of graphs containing a given site is exponential in $|g|$, so (\ref{(12)}) allows us to sum over collections of graphs connecting one site to another. We obtain an exponentially decaying connectivity function, demonstrating that resonant blocks do not percolate.

In order to relax simplifying assumption (b), we consider graphs 
with a significant fraction of flips appearing at previously flipped sites.
Such graphs no longer have the independent denominator property that was needed to obtain (\ref{(12)}). We resum collections of graphs for which the interval spanned by $g$ is smaller than $\frac{7}{8}|g|$. 
The shortened span implies an increased rate of decay with distance, so inductive bounds from previous steps are sufficient, and probabilistic estimates like (\ref{(12)}) are not needed. For the remaining ``straight'' graphs, most of the denominators are independent, and we obtain a bound similar to (\ref{(12)}) on the probability of resonance. To relax assumption (c), we need to incorporate the available $1/n!$ from the ad expansion into both sides of (\ref{(12)}) so as to allow summation over graphs of arbitrarily high order at a given scale $k$; see \cite{Imbrie2016} for details.

In order to completely diagonalize $H$, it is important to relax assumption (a) by considering graphs that involve resonant blocks. We know that resonant blocks are dilute, but their interactions could lead to long-range effects. 
A resonant block $B$ with diameter $d$ may have level spacings of order $2^{-d}$. In step $k$ we work with interactions of order $\gamma^{L_k}$, so we are able to resolve energy differences of order $\veps^{L_k}$. Therefore, we need to wait until $L_k$ becomes comparable to $d$ before considering interactions involving $B$. But larger interaction terms of order $\gamma^{L_j}$ from steps $j<k$ still connect $B$ to its immediate neighborhood. To get rid of these,
we define a fattened block $\bar{B}$ by adding an $L_k$-neighborhood to $B$.
Any graphs $g$ connecting $B$ to $\bar{B}^{\text{c}}$ (the complement of $\bar{B}$) must have $|g| \ge L_k$. We separate the interaction terms  of $H^{(k)} = H_0 + J^{(k)}$ into two parts: $J^{(k)\text{int}}$ (involving terms internal to $\bar{B}$) and $J^{(k)\text{ext}}$ (the rest). Let $O^{(k)}$ be the matrix that diagonalizes $H_0 + J^{(k)\text{int}}$, and use it to rotate $H^{(k)}$. In the new basis, we have an equivalent system with  $2^{|\bar{B}|}$ eigenstates of $\bar{B}$ interacting with $\bar{B}^{\text{c}}$ through terms of order $\gamma^{L_k}$ or smaller. In effect, $\bar{B}$ represents a ``fat'' site with a spin variable labeling the $2^{|\bar{B}|}$ basis states. We call this variable the metaspin for $\bar{B}$. Henceforth, eigenstates will be labeled by spin variables $\sigma_i = \pm 1$ away from resonant blocks, and metaspin variables in blocks $\bar{B}$. This is necessary because when a rotation is far from the identity, there is no natural way to assign eigenstates to the original basis states.

\begin{figure}[h]
\centering
\includegraphics{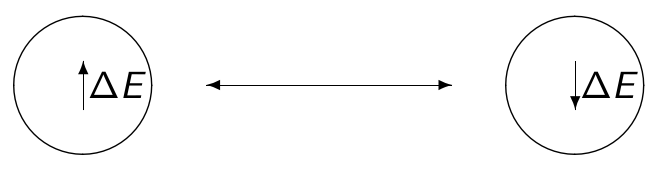}
\caption{A block-block resonance.\label{fig1}}
\end{figure}

In the new basis, we may rotate away interactions of order $\gamma^{L_k}$ and higher as discussed above. But to control the probability that a graph involving $\bar{B}$ is resonant, we need estimates like (\ref{(12)}), which depend on the finiteness of $\mathbb{E}|\Delta E|^{-s}$. This is where we need the assumption \textbf{LLA}$(\nu,C)$. By construction, the dependence of  $\Delta E$ for any level change in $\bar{B}$ depends on the state in $\bar{B}^{\text{c}}$ only out to a distance comparable to $L_k$. Therefore, we can approximate $\Delta E$ within an error $\gamma^{O(L_k)}$ by examining the equivalent transition for the Hamiltonian in an $L_k$-neighborhood of $\bar{B}$. Using \textbf{LLA}$(\nu,C)$ for a box of size $n$ comparable to $L_k$, we have that with high probability the minimum level spacing is at least $\tilde{C}^{-L_k}$ for some $\tilde{C}<\infty$. Furthermore, since all three terms in the Hamiltonian (\ref{(1)}) are random, the radial degree of freedom in the variables $\{h_i,\Gamma_i, J_i\}$ scales all energy differences equally. Hence as long as there is a minimum level spacing, we can control the probability that a transition energy $\Delta E$ is resonant with a given nearby transition -- even in the case of a block-block interaction as in Fig. \ref{fig1}. An energy difference $\Delta E$ of $\bar{B}$ moves linearly with the radial variable, so as before the fractional moment  $\mathbb{E}|\Delta E|^{-s}$ is bounded (in this case the bound is $\tilde{C}^{L_k}$). 

We see that a resonant region of size $d$ requires a buffer zone of width comparable to $d$ to enable it to become disentangled from neighboring degrees of freedom. Consequently, one should unite two such neighboring blocks into one if they are within a distance $d$. This leads to hierarchically organized, loosely connected resonant regions whose connectivity function decays as a stretched exponential. This idea is explored in a simple theory of the MBL transition in \cite{Zhang2016}. In dimensions greater than 1, the volume of $\bar{B}$ may be substantially larger than $d$. Consequently, the level spacing in $\bar{B}$ may become smaller than the size of the interaction terms, and our procedure breaks down.

Resonant blocks create gaps in decay for graphs passing through them. We need to ensure that there is minimal loss in decay from collections of resonant blocks on all scales. We want the usual metric in $\mathbb{Z}$ to be comparable to the one where blocks are contracted to points. Then graphs will decay exponentially in the usual metric, and those that traverse the buffer zone of $B$ are small enough to control the exponential number of states in $\bar{B}$. To this end, we use a more generous notion of connectivity wherein a block of volume $V$ (potentially much smaller than the diameter) is united with similar or larger blocks up to a distance $\exp(O(V^{1/2}))$ away. Then the separation between blocks grows much faster than their diameter, and the fraction of distance lost to blocks is summable and small. A similar construction was used in \cite{Frohlich1983}. This rule implies that block volumes scale up with their diameters at least as $O(|\log d|^2)$, which leads to a connectivity function that decays as
$(\gamma^\kappa)^{1+(\log|i-j|)^2}$, \textit{c.f.} Theorem \ref{thm:1}.

To complete the proof of Theorem \ref{thm:1}, let $k$ tend to infinity. This eliminates all off-diagonal entries of the Hamiltonian. All rotations are close to the identity, except in resonant blocks. As in the discussion of (\ref{(10)}), $\langle S^{\text{z}}_0 \rangle_\alpha$ is close to $\pm 1$ except for a set of measure $O(\veps) = O(\gamma^{1/20})$, and (\ref{(1.4)}) follows. An eigenstate correlation as in (\ref{(4)}) involves graphs extending from $i$ to $j$. If no more than half the distance from $i$ to $j$ is covered by resonant blocks, then we obtain exponential decay as in (\ref{(4)}). Otherwise, we have an event whose probability is dominated by the block connectivity function, which decays as indicated above (faster than any power of $1/|i-j|$).

In summary, we have described a multiscale approach to the problem of deforming the tensor product basis into the exact eigenfunctions. An assumption on level spacing allows probabilistic control over cases where perturbation theory is not well behaved; as a result we show that resonant regions fail to percolate, and MBL follows. We give convergent expansions for eigenvalues and eigenfunctions demonstrating closeness to the states from whence they came.


\begin{thebibliography}{30}%
\makeatletter
\providecommand \@ifxundefined [1]{%
 \@ifx{#1\undefined}
}%
\providecommand \@ifnum [1]{%
 \ifnum #1\expandafter \@firstoftwo
 \else \expandafter \@secondoftwo
 \fi
}%
\providecommand \@ifx [1]{%
 \ifx #1\expandafter \@firstoftwo
 \else \expandafter \@secondoftwo
 \fi
}%
\providecommand \natexlab [1]{#1}%
\providecommand \enquote  [1]{``#1''}%
\providecommand \bibnamefont  [1]{#1}%
\providecommand \bibfnamefont [1]{#1}%
\providecommand \citenamefont [1]{#1}%
\providecommand \href@noop [0]{\@secondoftwo}%
\providecommand \href [0]{\begingroup \@sanitize@url \@href}%
\providecommand \@href[1]{\@@startlink{#1}\@@href}%
\providecommand \@@href[1]{\endgroup#1\@@endlink}%
\providecommand \@sanitize@url [0]{\catcode `\\12\catcode `\$12\catcode
  `\&12\catcode `\#12\catcode `\^12\catcode `\_12\catcode `\%12\relax}%
\providecommand \@@startlink[1]{}%
\providecommand \@@endlink[0]{}%
\providecommand \url  [0]{\begingroup\@sanitize@url \@url }%
\providecommand \@url [1]{\endgroup\@href {#1}{\urlprefix }}%
\providecommand \urlprefix  [0]{URL }%
\providecommand \Eprint [0]{\href }%
\providecommand \doibase [0]{http://dx.doi.org/}%
\providecommand \selectlanguage [0]{\@gobble}%
\providecommand \bibinfo  [0]{\@secondoftwo}%
\providecommand \bibfield  [0]{\@secondoftwo}%
\providecommand \translation [1]{[#1]}%
\providecommand \BibitemOpen [0]{}%
\providecommand \bibitemStop [0]{}%
\providecommand \bibitemNoStop [0]{.\EOS\space}%
\providecommand \EOS [0]{\spacefactor3000\relax}%
\providecommand \BibitemShut  [1]{\csname bibitem#1\endcsname}%
\let\auto@bib@innerbib\@empty
\bibitem [{\citenamefont {Anderson}(1958)}]{Anderson1958}%
  \BibitemOpen
  \bibfield  {author} {\bibinfo {author} {\bibfnamefont {P.~W.}~\bibnamefont
  {Anderson}},\ }\href {\doibase 10.1103/PhysRev.109.1492} {\bibfield
  {journal} {\bibinfo  {journal} {Phys. Rev.}\ }\textbf {\bibinfo {volume}
  {109}},\ \bibinfo {pages} {1492} (\bibinfo {year} {1958})}\BibitemShut
  {NoStop}%
\bibitem [{\citenamefont {Fleishman}\ and\ \citenamefont
  {Anderson}(1980)}]{Fleishman1980}%
  \BibitemOpen
  \bibfield  {author} {\bibinfo {author} {\bibfnamefont {L.}~\bibnamefont
  {Fleishman}}\ and\ \bibinfo {author} {\bibfnamefont {P.~W.}~\bibnamefont
  {Anderson}},\ }\href {\doibase 10.1103/PhysRevB.21.2366} {\bibfield
  {journal} {\bibinfo  {journal} {Phys. Rev. B}\ }\textbf {\bibinfo {volume}
  {21}},\ \bibinfo {pages} {2366} (\bibinfo {year} {1980})}\BibitemShut
  {NoStop}%
\bibitem [{\citenamefont {Giamarchi}\ and\ \citenamefont
  {Schulz}(1987)}]{Giamarchi1987}%
  \BibitemOpen
  \bibfield  {author} {\bibinfo {author} {\bibfnamefont {T.}~\bibnamefont
  {Giamarchi}}\ and\ \bibinfo {author} {\bibfnamefont {H.}~\bibnamefont
  {Schulz}},\ }\href {\doibase 10.1209/0295-5075/3/12/007} {\bibfield
  {journal} {\bibinfo  {journal} {Europhys. Lett.}\ }\textbf {\bibinfo {volume}
  {3}},\ \bibinfo {pages} {1287} (\bibinfo {year} {1987})}\BibitemShut
  {NoStop}%
\bibitem [{\citenamefont {Altshuler}\ \emph {et~al.}(1997)\citenamefont
  {Altshuler}, \citenamefont {Gefen}, \citenamefont {Kamenev},\ and\
  \citenamefont {Levitov}}]{Altshuler1997}%
  \BibitemOpen
  \bibfield  {author} {\bibinfo {author} {\bibfnamefont {B.~L.}\ \bibnamefont
  {Altshuler}}, \bibinfo {author} {\bibfnamefont {Y.}~\bibnamefont {Gefen}},
  \bibinfo {author} {\bibfnamefont {A.}~\bibnamefont {Kamenev}}, \ and\
  \bibinfo {author} {\bibfnamefont {L.~S.}\ \bibnamefont {Levitov}},\ }\href
  {\doibase 10.1103/PhysRevLett.78.2803} {\bibfield  {journal} {\bibinfo
  {journal} {Phys. Rev. Lett.}\ }\textbf {\bibinfo {volume} {78}},\ \bibinfo
  {pages} {2803} (\bibinfo {year} {1997})}\BibitemShut {NoStop}%
\bibitem [{\citenamefont {Santos}\ \emph {et~al.}(2004)\citenamefont {Santos},
  \citenamefont {Rigolin},\ and\ \citenamefont {Escobar}}]{Santos2004}%
  \BibitemOpen
  \bibfield  {author} {\bibinfo {author} {\bibfnamefont {L.~F.}~\bibnamefont
  {Santos}}, \bibinfo {author} {\bibfnamefont {G.}~\bibnamefont {Rigolin}}, \
  and\ \bibinfo {author} {\bibfnamefont {C.~O.}~\bibnamefont {Escobar}},\ }\href
  {\doibase 10.1103/PhysRevA.69.042304} {\bibfield  {journal} {\bibinfo
  {journal} {Phys. Rev. A}\ }\textbf {\bibinfo {volume} {69}},\ \bibinfo
  {pages} {042304} (\bibinfo {year} {2004})}\BibitemShut {NoStop}%
\bibitem [{\citenamefont {Gornyi}\ \emph {et~al.}(2005)\citenamefont {Gornyi},
  \citenamefont {Mirlin},\ and\ \citenamefont {Polyakov}}]{Gornyi2005}%
  \BibitemOpen
  \bibfield  {author} {\bibinfo {author} {\bibfnamefont {I.~V.}~\bibnamefont
  {Gornyi}}, \bibinfo {author} {\bibfnamefont {A.~D.}~\bibnamefont {Mirlin}}, \
  and\ \bibinfo {author} {\bibfnamefont {D.~G.}~\bibnamefont {Polyakov}},\ }\href
  {\doibase 10.1103/PhysRevLett.95.206603} {\bibfield  {journal} {\bibinfo
  {journal} {Phys. Rev. Lett.}\ }\textbf {\bibinfo {volume} {95}},\ \bibinfo
  {pages} {206603} (\bibinfo {year} {2005})}\BibitemShut {NoStop}%
\bibitem [{\citenamefont {Basko}\ \emph {et~al.}(2006)\citenamefont {Basko},
  \citenamefont {Aleiner},\ and\ \citenamefont {Altshuler}}]{Basko2006}%
  \BibitemOpen
  \bibfield  {author} {\bibinfo {author} {\bibfnamefont {D.~M.}\ \bibnamefont
  {Basko}}, \bibinfo {author} {\bibfnamefont {I.~L.}\ \bibnamefont {Aleiner}},
  \ and\ \bibinfo {author} {\bibfnamefont {B.~L.}\ \bibnamefont {Altshuler}},\
  }\href {\doibase 10.1016/j.aop.2005.11.014} {\bibfield  {journal} {\bibinfo
  {journal} {Ann. Phys. (N. Y).}\ }\textbf {\bibinfo {volume} {321}},\ \bibinfo
  {pages} {1126} (\bibinfo {year} {2006})}\BibitemShut {NoStop}%
\bibitem [{\citenamefont {{\v{Z}}nidari{\v{c}}}\ \emph
  {et~al.}(2008)\citenamefont {{\v{Z}}nidari{\v{c}}}, \citenamefont {Prosen},\
  and\ \citenamefont {Prelov{\v{s}}ek}}]{Znidaric2008}%
  \BibitemOpen
  \bibfield  {author} {\bibinfo {author} {\bibfnamefont {M.}~\bibnamefont
  {{\v{Z}}nidari{\v{c}}}}, \bibinfo {author} {\bibfnamefont {T.}~\bibnamefont
  {Prosen}}, \ and\ \bibinfo {author} {\bibfnamefont {P.}~\bibnamefont
  {Prelov{\v{s}}ek}},\ }\href {\doibase 10.1103/PhysRevB.77.064426} {\bibfield
  {journal} {\bibinfo  {journal} {Phys. Rev. B}\ }\textbf {\bibinfo {volume}
  {77}},\ \bibinfo {pages} {064426} (\bibinfo {year} {2008})}\BibitemShut
  {NoStop}%
\bibitem [{\citenamefont {Oganesyan}\ and\ \citenamefont
  {Huse}(2007)}]{Oganesyan2007}%
  \BibitemOpen
  \bibfield  {author} {\bibinfo {author} {\bibfnamefont {V.}~\bibnamefont
  {Oganesyan}}\ and\ \bibinfo {author} {\bibfnamefont {D.~A.}\ \bibnamefont
  {Huse}},\ }\href {\doibase 10.1103/PhysRevB.75.155111} {\bibfield  {journal}
  {\bibinfo  {journal} {Phys. Rev. B}\ }\textbf {\bibinfo {volume} {75}},\
  \bibinfo {pages} {155111} (\bibinfo {year} {2007})}\BibitemShut {NoStop}%
\bibitem [{\citenamefont {Pal}\ and\ \citenamefont {Huse}(2010)}]{Pal2010}%
  \BibitemOpen
  \bibfield  {author} {\bibinfo {author} {\bibfnamefont {A.}~\bibnamefont
  {Pal}}\ and\ \bibinfo {author} {\bibfnamefont {D.~A.}~\bibnamefont {Huse}},\
  }\href {\doibase 10.1103/PhysRevB.82.174411} {\bibfield  {journal} {\bibinfo
  {journal} {Phys. Rev. B}\ }\textbf {\bibinfo {volume} {82}},\ \bibinfo
  {pages} {174411} (\bibinfo {year} {2010})}\BibitemShut {NoStop}%
\bibitem [{\citenamefont {Bauer}\ and\ \citenamefont
  {Nayak}(2013)}]{Bauer2013}%
  \BibitemOpen
  \bibfield  {author} {\bibinfo {author} {\bibfnamefont {B.}~\bibnamefont
  {Bauer}}\ and\ \bibinfo {author} {\bibfnamefont {C.}~\bibnamefont {Nayak}},\
  }\href {\doibase 10.1088/1742-5468/2013/09/P09005} {\bibfield  {journal}
  {\bibinfo  {journal} {J. Stat. Mech. Theory Exp.}\ }\textbf {\bibinfo
  {volume} {2013}},\ \bibinfo {pages} {P09005} (\bibinfo {year}
  {2013})}\BibitemShut {NoStop}%
\bibitem [{\citenamefont {Nandkishore}\ and\ \citenamefont
  {Huse}(2015)}]{Nandkishore2015}%
  \BibitemOpen
  \bibfield  {author} {\bibinfo {author} {\bibfnamefont {R.}~\bibnamefont
  {Nandkishore}}\ and\ \bibinfo {author} {\bibfnamefont {D.~A.}\ \bibnamefont
  {Huse}},\ }\href {\doibase 10.1146/annurev-conmatphys-031214-014726}
  {\bibfield  {journal} {\bibinfo  {journal} {Annu. Rev. Condens. Matter
  Phys.}\ }\textbf {\bibinfo {volume} {6}},\ \bibinfo {pages} {15} (\bibinfo
  {year} {2015})}\BibitemShut {NoStop}%
\bibitem [{\citenamefont {Griffiths}(1969)}]{Griffiths1969}%
  \BibitemOpen
  \bibfield  {author} {\bibinfo {author} {\bibfnamefont {R.~B.}\ \bibnamefont
  {Griffiths}},\ }\href {\doibase 10.1103/PhysRevLett.23.17} {\bibfield
  {journal} {\bibinfo  {journal} {Phys. Rev. Lett.}\ }\textbf {\bibinfo
  {volume} {23}},\ \bibinfo {pages} {17} (\bibinfo {year} {1969})}\BibitemShut
  {NoStop}%
\bibitem [{\citenamefont {Hamza}\ \emph {et~al.}(2012)\citenamefont {Hamza},
  \citenamefont {Sims},\ and\ \citenamefont {Stolz}}]{Hamza2012a}%
  \BibitemOpen
  \bibfield  {author} {\bibinfo {author} {\bibfnamefont {E.}~\bibnamefont
  {Hamza}}, \bibinfo {author} {\bibfnamefont {R.}~\bibnamefont {Sims}}, \ and\
  \bibinfo {author} {\bibfnamefont {G.}~\bibnamefont {Stolz}},\ }\href
  {\doibase 10.1007/s00220-012-1544-6} {\bibfield  {journal} {\bibinfo
  {journal} {Commun. Math. Phys.}\ }\textbf {\bibinfo {volume} {315}},\
  \bibinfo {pages} {215} (\bibinfo {year} {2012})}\BibitemShut {NoStop}%
\bibitem [{\citenamefont {Mastropietro}(2015)}]{Mastropietro2015}%
  \BibitemOpen
  \bibfield  {author} {\bibinfo {author} {\bibfnamefont {V.}~\bibnamefont
  {Mastropietro}},\ }\href {\doibase 10.1103/PhysRevLett.115.180401} {\bibfield
   {journal} {\bibinfo  {journal} {Phys. Rev. Lett.}\ }\textbf {\bibinfo
  {volume} {115}},\ \bibinfo {pages} {180401} (\bibinfo {year}
  {2015})}\BibitemShut {NoStop}%
\bibitem [{\citenamefont {Imbrie}\ and\ \citenamefont
  {Mavi}(2016)}]{Imbrie2016a}%
  \BibitemOpen
  \bibfield  {author} {\bibinfo {author} {\bibfnamefont {J.~Z.}\ \bibnamefont
  {Imbrie}}\ and\ \bibinfo {author} {\bibfnamefont {R.}~\bibnamefont {Mavi}},\
  }\href {\doibase 10.1007/s10955-016-1461-8} {\bibfield  {journal} {\bibinfo
  {journal} {J. Stat. Phys.}\ }\textbf {\bibinfo {volume} {162}},\ \bibinfo
  {pages} {1451} (\bibinfo {year} {2016})}\BibitemShut {NoStop}%
\bibitem [{\citenamefont {Deutsch}(1991)}]{Deutsch1991}%
  \BibitemOpen
  \bibfield  {author} {\bibinfo {author} {\bibfnamefont {J.~M.}~\bibnamefont
  {Deutsch}},\ }\href {\doibase 10.1103/PhysRevA.43.2046} {\bibfield  {journal}
  {\bibinfo  {journal} {Phys. Rev. A}\ }\textbf {\bibinfo {volume} {43}},\
  \bibinfo {pages} {2046} (\bibinfo {year} {1991})}\BibitemShut {NoStop}%
\bibitem [{\citenamefont {Srednicki}(1994)}]{Srednicki1994}%
  \BibitemOpen
  \bibfield  {author} {\bibinfo {author} {\bibfnamefont {M.}~\bibnamefont
  {Srednicki}},\ }\href {\doibase 10.1103/PhysRevE.50.888} {\bibfield
  {journal} {\bibinfo  {journal} {Phys. Rev. E}\ }\textbf {\bibinfo {volume}
  {50}},\ \bibinfo {pages} {888} (\bibinfo {year} {1994})}\BibitemShut
  {NoStop}%
\bibitem [{\citenamefont {Rigol}\ \emph {et~al.}(2008)\citenamefont {Rigol},
  \citenamefont {Dunjko},\ and\ \citenamefont {Olshanii}}]{Rigol2008}%
  \BibitemOpen
  \bibfield  {author} {\bibinfo {author} {\bibfnamefont {M.}~\bibnamefont
  {Rigol}}, \bibinfo {author} {\bibfnamefont {V.}~\bibnamefont {Dunjko}}, \
  and\ \bibinfo {author} {\bibfnamefont {M.}~\bibnamefont {Olshanii}},\ }\href
  {\doibase 10.1038/nature06838} {\bibfield  {journal} {\bibinfo  {journal}
  {Nature}\ }\textbf {\bibinfo {volume} {452}},\ \bibinfo {pages} {854}
  (\bibinfo {year} {2008})}\BibitemShut {NoStop}%
\bibitem [{\citenamefont {Imbrie}(2016)}]{Imbrie2016}%
  \BibitemOpen
  \bibfield  {author} {\bibinfo {author} {\bibfnamefont {J.~Z.}\ \bibnamefont
  {Imbrie}},\ }\href {\doibase 10.1007/s10955-016-1508-x} {\bibfield  {journal}
  {\bibinfo  {journal} {J. Stat. Phys.}\ } \textbf {\bibinfo {volume} {163}},\ \bibinfo {pages} {998}
  (\bibinfo {year} {2016})}\BibitemShut {NoStop}%
\bibitem [{\citenamefont {Huse}\ \emph {et~al.}(2014)\citenamefont {Huse},
  \citenamefont {Nandkishore},\ and\ \citenamefont {Oganesyan}}]{Huse2014}%
  \BibitemOpen
  \bibfield  {author} {\bibinfo {author} {\bibfnamefont {D.~A.}\ \bibnamefont
  {Huse}}, \bibinfo {author} {\bibfnamefont {R.}~\bibnamefont {Nandkishore}}, \
  and\ \bibinfo {author} {\bibfnamefont {V.}~\bibnamefont {Oganesyan}},\ }\href
  {\doibase 10.1103/PhysRevB.90.174202} {\bibfield  {journal} {\bibinfo
  {journal} {Phys. Rev. B}\ }\textbf {\bibinfo {volume} {90}},\ \bibinfo
  {pages} {174202} (\bibinfo {year} {2014})}\BibitemShut {NoStop}%
\bibitem [{\citenamefont {Serbyn}\ \emph {et~al.}(2013)\citenamefont {Serbyn},
  \citenamefont {Papi{\'{c}}},\ and\ \citenamefont {Abanin}}]{Serbyn2013}%
  \BibitemOpen
  \bibfield  {author} {\bibinfo {author} {\bibfnamefont {M.}~\bibnamefont
  {Serbyn}}, \bibinfo {author} {\bibfnamefont {Z.}~\bibnamefont {Papi{\'{c}}}},
  \ and\ \bibinfo {author} {\bibfnamefont {D.~A.}\ \bibnamefont {Abanin}},\
  }\href {\doibase 10.1103/PhysRevLett.111.127201} {\bibfield  {journal}
  {\bibinfo  {journal} {Phys. Rev. Lett.}\ }\textbf {\bibinfo {volume} {111}},\
  \bibinfo {pages} {127201} (\bibinfo {year} {2013})}\BibitemShut {NoStop}%
\bibitem [{\citenamefont {Ros}\ \emph {et~al.}(2015)\citenamefont {Ros},
  \citenamefont {M{\"{u}}ller},\ and\ \citenamefont {Scardicchio}}]{Ros2015}%
  \BibitemOpen
  \bibfield  {author} {\bibinfo {author} {\bibfnamefont {V.}~\bibnamefont
  {Ros}}, \bibinfo {author} {\bibfnamefont {M.}~\bibnamefont {M{\"{u}}ller}}, \
  and\ \bibinfo {author} {\bibfnamefont {A.}~\bibnamefont {Scardicchio}},\
  }\href {\doibase 10.1016/j.nuclphysb.2014.12.014} {\bibfield  {journal}
  {\bibinfo  {journal} {Nucl. Phys. B}\ }\textbf {\bibinfo {volume} {891}},\
  \bibinfo {pages} {420} (\bibinfo {year} {2015})}\BibitemShut {NoStop}%
\bibitem [{\citenamefont {Imbrie}(2015)}]{Imbrie2015}%
  \BibitemOpen
  \bibfield  {author} {\bibinfo {author} {\bibfnamefont {J.~Z.}\ \bibnamefont
  {Imbrie}},\ }\href {\doibase 10.1007/s00220-015-2522-6} {\bibfield  {journal}
  {\bibinfo  {journal} {Commun. Math. Phys.}\ }\textbf {\bibinfo {volume}
  {341}},\ \bibinfo {pages} {491} (\bibinfo {year} {2015})}\BibitemShut
  {NoStop}%
\bibitem [{\citenamefont {Bellissard}\ \emph {et~al.}(1983)\citenamefont
  {Bellissard}, \citenamefont {Lima},\ and\ \citenamefont
  {Testard}}]{Bellissard1983b}%
  \BibitemOpen
  \bibfield  {author} {\bibinfo {author} {\bibfnamefont {J.}~\bibnamefont
  {Bellissard}}, \bibinfo {author} {\bibfnamefont {R.}~\bibnamefont {Lima}}, \
  and\ \bibinfo {author} {\bibfnamefont {D.}~\bibnamefont {Testard}},\ }\href
  {http://projecteuclid.org/euclid.cmp/1103922281} {\bibfield  {journal}
  {\bibinfo  {journal} {Commun. Math. Phys.}\ }\textbf {\bibinfo {volume}
  {88}},\ \bibinfo {pages} {207} (\bibinfo {year} {1983})}\BibitemShut
  {NoStop}%
\bibitem [{\citenamefont {Chulaevsky}\ and\ \citenamefont
  {Sinai}(1991)}]{Chulaevsky1991}%
  \BibitemOpen
  \bibfield  {author} {\bibinfo {author} {\bibfnamefont {V.}~\bibnamefont
  {Chulaevsky}}\ and\ \bibinfo {author} {\bibfnamefont {Y.}~\bibnamefont
  {Sinai}},\ }\href {\doibase 10.1142/S0129055X91000096} {\bibfield  {journal}
  {\bibinfo  {journal} {Rev. Math. Phys.}\ }\textbf {\bibinfo {volume} {03}},\
  \bibinfo {pages} {241} (\bibinfo {year} {1991})}\BibitemShut {NoStop}%
\bibitem [{\citenamefont {Eliasson}(1997)}]{Eliasson1997}%
  \BibitemOpen
  \bibfield  {author} {\bibinfo {author} {\bibfnamefont {L.}~\bibnamefont
  {Eliasson}},\ }\href {\doibase 10.1007/BF02392742} {\bibfield  {journal}
  {\bibinfo  {journal} {Acta Math.}\ }\textbf {\bibinfo {volume} {179}},\
  \bibinfo {pages} {153} (\bibinfo {year} {1997})}\BibitemShut {NoStop}%
\bibitem [{\citenamefont {Datta}\ \emph {et~al.}(1996)\citenamefont {Datta},
  \citenamefont {Fern{\'{a}}ndez},\ and\ \citenamefont
  {Fr{\"{o}}hlich}}]{Datta1996}%
  \BibitemOpen
  \bibfield  {author} {\bibinfo {author} {\bibfnamefont {N.}~\bibnamefont
  {Datta}}, \bibinfo {author} {\bibfnamefont {R.}~\bibnamefont
  {Fern{\'{a}}ndez}}, \ and\ \bibinfo {author} {\bibfnamefont {J.}~\bibnamefont
  {Fr{\"{o}}hlich}},\ }\href {\doibase 10.1007/BF02179651} {\bibfield
  {journal} {\bibinfo  {journal} {J. Stat. Phys.}\ }\textbf {\bibinfo {volume}
  {84}},\ \bibinfo {pages} {455} (\bibinfo {year} {1996})}\BibitemShut
  {NoStop}%
\bibitem [{\citenamefont {Zhang}\ \emph {et~al.}()\citenamefont {Zhang},
  \citenamefont {Zhao}, \citenamefont {Devakul},\ and\ \citenamefont
  {Huse}}]{Zhang2016}%
  \BibitemOpen
  \bibfield  {author} {\bibinfo {author} {\bibfnamefont {L.}~\bibnamefont
  {Zhang}}, \bibinfo {author} {\bibfnamefont {B.}~\bibnamefont {Zhao}},
  \bibinfo {author} {\bibfnamefont {T.}~\bibnamefont {Devakul}}, \ and\
  \bibinfo {author} {\bibfnamefont {D.~A.}\ \bibnamefont {Huse}},\ }\href
  {http://arxiv.org/abs/1603.02296} {\ }\Eprint
  {http://arxiv.org/abs/1603.02296} {arXiv:1603.02296} \BibitemShut {NoStop}%
\bibitem [{\citenamefont {Fr{\"{o}}hlich}\ and\ \citenamefont
  {Spencer}(1983)}]{Frohlich1983}%
  \BibitemOpen
  \bibfield  {author} {\bibinfo {author} {\bibfnamefont {J.}~\bibnamefont
  {Fr{\"{o}}hlich}}\ and\ \bibinfo {author} {\bibfnamefont {T.}~\bibnamefont
  {Spencer}},\ }\href {http://projecteuclid.org/euclid.cmp/1103922279}
  {\bibfield  {journal} {\bibinfo  {journal} {Commun. Math. Phys.}\ }\textbf
  {\bibinfo {volume} {88}},\ \bibinfo {pages} {151} (\bibinfo {year}
  {1983})}\BibitemShut {NoStop}%
\end{thebibliography}
\end{document}